\documentclass[10pt]{article}

\usepackage{amsmath}
\usepackage{amssymb}
\usepackage{mathrsfs}
\usepackage[english]{babel}
\usepackage{hyperref}
\usepackage{array,graphicx}
\usepackage{subcaption}
\usepackage{booktabs} 
\usepackage{caption}
\usepackage{float}
\usepackage{multicol}
\usepackage{colortbl}
\usepackage{fancyhdr}
\usepackage{multirow}
\usepackage{changepage}
\usepackage{makecell}
\usepackage[dvipsnames]{xcolor}
\usepackage[a4paper]{geometry}
\usepackage{pifont}
\newcommand*\rot{\rotatebox{90}}
\newcommand*\OK{\ding{51}}

\newcommand{\revision}[1]{{\color{black}#1}}

\newtheorem{defin}{\textit{Definition}}[section]
\newtheorem{theor}{\textit{Theorem}}[section]

\title{Benchmarking tools for \textit{a priori} identifiability analysis}
\author{Xabier Rey Barreiro$^1$, Alejandro F. Villaverde$^{1,2}$\\
{\footnotesize $^1$ Universidade de Vigo, Department of Systems Engineering \& Control, 36310 Vigo, Galicia, Spain}\\
{\footnotesize $^2$ CITMAga, 15782 Santiago de Compostela, Galicia, Spain}\\
{\footnotesize emails: xabier.rey@uvigo.gal, afvillaverde@uvigo.gal}
}
\date{}

\begin{document}

\maketitle

\abstract{
The structural identifiability and the observability of a model determine the possibility of inferring its parameters and states by observing its outputs. These properties should be analysed before attempting to calibrate a model. Unfortunately, such \textit{a priori} analysis can be challenging, since it requires symbolic calculations that often have a high computational cost. In recent years a number of software tools have been developed for this task, 
mostly in the systems biology community but also in other disciplines. These tools have vastly different features and capabilities, and a critical assessment of their performance is still lacking. Here we present a comprehensive study of the computational resources available for analysing structural identifiability. We consider 12 software tools developed in 7 programming languages (Matlab, Maple, Mathematica, Julia, Python, Reduce, and Maxima), and evaluate their performance using a set of 25 case studies created from 21 models. Our results reveal their strengths and weaknesses, provide guidelines for choosing the most appropriate tool for a given problem, and highlight opportunities for future developments.
}

\section{Introduction}

Mathematical modelling is an essential tool for describing the dynamics of natural and artificial systems. In systems biology, model dynamics are often given by nonlinear ordinary differential equations (ODEs). As these models typically have unknown parameters, it is necessary to determine their values by fitting the model to experimental data. This task, known as model calibration or parameter estimation \cite{villaverde2022protocol}, can only be performed successfully if the model is identifiable. If some of the unknown parameters are unidentifiable, their values cannot be determined by measuring the model output. Lack of identifiability may lead to inaccurate inferences of mechanistically meaningful parameters, as well as to the inability to make correct predictions of certain variables. In order to avoid such errors it is important to detect unidentifiability and to identify its sources \cite{munoz2018or,janzen2016parameter,eisenberg2017confidence}. 

It is common to distinguish between structural and practical identifiability \cite{wieland2021structural}.
Structural identifiability is a theoretical property that is fully determined by the model equations, that is, it depends on the system dynamics, the measurable outputs, and the admissible inputs \cite{bellman1970structural}. It is also called \textit{a priori} identifiability, since it can be tested before performing experiments and collecting data. We note that some authors consider \textit{a priori} identifiability as a particular type of structural identifiability \cite{wieland2021structural}, while others use both terms interchangeably \cite{anstett2020priori}; in this paper we adopt the latter terminology.
A related property is observability, which is the possibility of inferring the internal state of a system from observations of its outputs. By considering model parameters as constant state variables, \textit{a priori} identifiability can be recast as a particular case of observability \cite{tunali1987new}. 
When a model has structural unidentifiabilities, it is due to inadequacies in its equations, such as the existence of Lie symmetries \cite{sedoglavic2007reduction,anguelova2012minimal,merkt2015higher}. In order to remove a structural unidentifiability it is necessary to modify the model equations, for example by reparameterizing the ODEs or by enlarging the output function.
In contrast, practical unidentifiabilities are due to lack of sufficiently informative experimental data, and they can be overcome by using alternative or additional datasets for model calibration \cite{wolkenhauer2008parameter,apgar2010sloppy}.
Structural (\textit{a priori}) identifiability is a prerequisite for practical (\textit{a posteriori}) identifiability. It is essential to analyse this property before attempting to perform parameter estimation, since the ways of overcoming practical and structural unidentifiability are different.
A further distinction can be made within structural identifiability, giving rise to two different properties: structural \textit{local} identifiability (SLI) and structural \textit{global} identifiability (SGI). A parameter that has the SLI property can be uniquely inferred in a neighbourhood of its nominal value, but a finite number of indistinguishable solutions may exist in the parameter space. In contrast, a parameter with the SGI property has a unique solution in the whole parameter space.

Unfortunately, the \textit{a priori} analysis of identifiability (and observability) is mathematically involved, and it can be computationally challenging. It requires symbolic computations that quickly become very costly even for models of moderate size. Hence, a number of methodologies have been developed for its study, as well as specialised software tools. 
Two key papers \cite{miao2011identifiability,chis2011structural} provided an overview of the state of the art in 2011: Miao \textit{et al} \cite{miao2011identifiability} reviewed the theoretical foundations of practical and structural identifiability analysis methods, and Chis \textit{et al} \cite{chis2011structural} performed a computational comparison of structural identifiability algorithms. At that time, only two software toolboxes were publicly available for this task: DAISY \cite{bellu2007daisy} and GenSSI \cite{chics2011genssi}; hence, Chis \textit{et al} implemented a number of other approaches themselves. In 2013, Raue \textit{et al} compared DAISY with two other identifiability analysis tools that had been recently presented \cite{raue2014comparison}: the Exact Arithmetic Rank (EAR), implemented in Mathematica \cite{karlsson2012efficient}, and the Profile Likelihood (PL), which is a numerical technique for \textit{a posteriori} analysis \cite{raue2009structural}.

Since the publication of \cite{raue2014comparison}, a significant number of software tools for structural identifiability analysis have been presented, including the web app COMBOS \cite{meshkat2014finding}, the Matlab toolboxes STRIKE-GOLDD \cite{villaverde2016structural}, GenSSI2 \cite{ligon2018genssi}, ORC-DF \revision{\cite{maes2019observability}, and rational ORC-DF or RORC-DF} \cite{shi2022efficient}, the Maple toolboxes SIAN \cite{hong20199sian} and ObservabilityTest (based on \cite{sedoglavic2002probabilistic}), the Julia packages StructuralIdentifiability \cite{dong2022differential} and SIAN \cite{hong20199sian}, and the Python tool StrikePy \cite{rey2022strikepy}. However, an assessment of their relative strengths and witnesses is currently lacking. Given their different theoretical foundations, capabilities, and computational performances, there is a clear need for their critical analysis and comparison. Some results in this direction were presented in \cite{hong20199sian}, where the performance of four tools for structural global identifiability analysis (DAISY, COMBOS, GenSSI, and SIAN) was compared using six case studies.

In this article we address this need by performing a thorough comparison of the software tools currently available for analysing structural identifiability and observability. We consider symbolic computation methods, which perform said analyses \textit{a priori}. We do not consider numerical approaches, such as the aforementioned PL \cite{raue2009structural} or sensitivity-based methods \cite{stigter2015fast}, which perform \textit{a posteriori} analyses of identifiability and can complement the techniques reviewed here \cite{wieland2021structural}. Thus, we have evaluated twelve different tools, available in seven different environments: Matlab, Maple, Mathematica, Julia, Python, Reduce, and web-based applications. To evaluate their performance we use a total of 25 variants of 21 basic models, of different sizes and complexities, taken from the systems biology literature. We discuss the strengths and weaknesses of each tool, and provide guidelines for choosing the most appropriate tool for a given problem. Our results represent the most comprehensive, up-to-date study of the available tools for structural identifiability analysis.

\section{Methods}

\subsection{Background on structural identifiability and observability}

We consider dynamic models described by ordinary differential equations in state space form:

\begin{equation}
\Sigma = 
\label{eqn:SD}
\begin{cases}
    \Dot{x} \thinspace = \thinspace f(t,x(t),u(t),\theta,w(t)),\\
    y(t) \thinspace = \thinspace h(x(t),u(t),\theta,w(t)),\\
    x(0) \thinspace=\thinspace x^0(\theta)
\end{cases}
\end{equation}
where $x(t) \thinspace \in \mathbb R^{n}$ is a vector of state variables, $y(t) \thinspace \in \mathbb R^{m}$ is a vector of outputs or measurements, $u(t) \thinspace \in \mathbb R^{q}$ is the vector of known inputs, $w(t) \thinspace \in \mathbb R^{q_w}$ is the vector of unknown inputs, and $\theta \thinspace \in \mathbb R^{p}$ is the unknown parameter vector. Initial conditions may be functions of unknown parameters, or generic unknown values. We write individual parameters and state variables with subindices (i.e. $\theta_i$, $x_i$), and we denote as $y(t,\theta^*)$ the output of a model $\Sigma$ for a specific parameter vector $\theta^*$.

\subsubsection{Definitions}

Many definitions of \textit{a priori} identifiability can be found in the literature. They describe similar properties with subtle differences among them. For a detailed account of said definitions and their nuances we refer the interested reader to \cite{anstett2020priori}. In what follows we provide only brief descriptions of these concepts, which we attempt to keep as simple as possible.

Roughly speaking, a dynamic model of the form \eqref{eqn:SD} is said to be \textit{observable} if its current state vector $x(t)$ can be determined from knowledge of the future values of the output $y(t)$ and input functions $u(t)$ in finite time.
Likewise, it is \textit{identifiable} if its parameter vector $\theta$ can be determined from the output $y(t)$ and input functions $u(t)$ in finite time.
It is common to distinguish between local and global identifiability. 

\begin{defin} Structural Local Identifiability: a parameter $\theta_i$ of a dynamic model $\Sigma$ is structurally locally identifiable (SLI) if, for almost all possible parameter vectors and almost all initial conditions, there is a neighbourhood $\mathcal{N}(\theta^*)$ in which the equality $y(t,\tilde{\theta})=y(t,\theta^*)$ holds if and only if $\tilde{\theta}_i=\theta_i^*$.
\end{defin}

\begin{defin} Structural Global Identifiability: a parameter $\theta_i$ of a dynamic model $\Sigma$ is structurally globally identifiable (SGI) if, for almost all possible parameter vectors and almost all initial conditions, the equality $y(t,\tilde{\theta})=y(t,\theta^*)$ holds if and only if $\tilde{\theta}_i=\theta_i^*$.
\end{defin}
Note that SGI parameters are also SLI.
For a SLI parameter there is a finite number of solutions, while for a SGI parameter there is a unique solution.
If the above conditions do not hold, the parameter is \textit{structurally unidentifiable} (SU). A model is said to be SGI (respectively, SLI) if all its parameters are SGI (resp., at least SLI). If it has at least one SU parameter, the model is called SU.

Likewise, we could provide local and global definitions of the observability of the system states. Nevertheless, the theory of observability of nonlinear systems was originally developed in a differential geometric framework \cite{hermann1977nonlinear} as a local property, and it is therefore common to consider observability only from a local point of view. Hence we use the following definition:
\begin{defin} Observability:
a state variable $x_i(\tau)$ is observable if, for almost all possible parameter vectors and almost all initial conditions, there is a neighbourhood $\mathcal{N}(\theta^*)$ in which the equality $y(t,\tilde{x}(\tau)) = y(t,{x}^*(\tau))$ holds if and only if $\tilde{x}_i(\tau)=x_i^*(\tau)$.
\end{defin}

\subsubsection{The differential geometry approach}

Structural \textit{local} identifiability can be analysed with a differential geometric approach, which is based on evaluating the Observability Rank Condition (ORC). Before defining the ORC we need to define the property that it assesses, i.e. the \textit{local weak observability}: 

\begin{defin} Local weak observability \cite{hermann1977nonlinear}. 
Let $U$ be an open subset in $\mathbb R^n$, and let indistinguishability be an equivalence relation on $\mathbb R^n$. 
We denote as $I(x_0,U)$ all points $x_i \in U$ that are indistinguishable from $x_0$.
The system $\Sigma$ is \textit{locally weakly observable} at $x_0$ if $I(x_0,V) = x_0$ for every open neighbourhood $V$ of $x_0$ contained in $U$.
\end{defin}

According to the above definition, $\Sigma$ is locally weakly observable if it is possible to distinguish each state vector from its neighbours.
Local weak observability is a property of the states; however, it can also be applied to the parameters by considering them as constant state variables, i.e. with zero dynamics \cite{tunali1987new}. In this view, a SLI parameter is a weakly locally observable state and this approach can be used to test whether a parameter is SLI. 

Before defining the ORC we need a few more mathematical preliminaries. Let $L_v(\phi)(x) := <d\phi,v>$ denote the differentiation of an infinitely differentiable function $\phi$ on $\mathbb R^n$ by a vector field $v$ on $\mathbb R^n$ \cite{anguelova2004nonlinear}, where $d\phi$ is the gradient of $\phi$ and <> the scalar product. We denote by $\Phi(t,x)$ the flow of a vector field $v$ on $\mathbb R^n$. The Taylor series of $\phi(\Phi(t,x))$ with respect to $t$ are called Lie series and are given by: 

$$\phi(\Phi(t,x)) = \sum^{\infty}_{k=0} \dfrac{t^k}{k!} L^k_v (\phi)(x)$$

Let $\varrho$ denote the space spanned by $L_f^qh_i$ at $x_0$ for $q\geq 0$ and $i=1,..,m$, for all vector fields $f(x,u)$. 
The space spanned by the gradients of the elements of $\varrho$ is defined by $d\varrho = \text{span}_{\mathbb R_x} \{ d\phi: \hspace{3mm} \phi \in \varrho \}$, where $\mathbb R_x$ indicates the field of meromorphic functions on $\mathbb R^n$. 
In certain contexts, $d\varrho$ is known as the observability matrix, $O(x)$, and its dimension determines the local weak observability property. 
Thus, the observability -- and therefore the structural local identifiability -- of a model can be tested with the following theorem:

\begin{theor}
\label{ORC}
Observability Rank Condition (ORC) \cite{hermann1977nonlinear}: if the system $\Sigma$ (\ref{eqn:SD}) satisfies $rank(O(x_0)) = n$, then it is locally weakly observable around $x_0$.
\end{theor}

\subsubsection{The differential algebra approach}

Structural \textit{global} identifiability can be tested with a differential algebra approach. It relies on finding algebraic equations that relate the model parameters with the inputs and outputs \cite{ljung1994global}.
Importantly, this approach introduces a restriction on the class of systems that can be analysed: instead of being applicable to general nonlinear ODE systems of the form \eqref{eqn:SD}, it requires that the ODE functions are rational. The same restriction is shared by other methods, as will be detailed in Section \ref{sec:tools}.

\begin{defin} Let $C^N_u[0,T]$ denote the function space expanded by all inputs on $[0,T]$ which have continuous derivatives up to the order N. The system $\Sigma$ is said to be algebraically identifiable if a meromorphic function exists,
$$
\Phi = \phi(\theta,u,\Dot{u}, ..., u^{(k)}, y, \Dot{y}, ..., y^{(k)}),\hspace{2mm} \Phi \in \mathbb R^p,
$$
which can be derived from a finite number of differentiation or algebraic calculation steps, so that the following equations, $\Phi = 0$ and $\text{det}\dfrac{\partial \Phi}{\partial \theta} \neq 0$, hold in the time range $[0, T]$, for all $(\theta,x_0,u)$ in an open and dense subset of $\Theta \times M \times C_u^N[0,T]$. Here, $k$ denotes a positive integer and $\Dot{u},...,u^{(k)}$ are the derivatives of u, and $\Dot{y}, ..., y^{(k)}$ the derivatives of y.
\end{defin}

Differential algebra methods replace the equations \ref{eqn:SD} of the system $\Sigma$ by a set of $m+n$ polynomial differential equations that depend only on the variables $(y,u)$, i.e., they rewrite $\Sigma$ in implicit form \cite{saccomani2001new}. These differential equations, known as the \textit{characteristic set}, preserve the dynamics of the model output while eliminating the state variables from the equations; they are obtained by applying Ritt's differential algebra \cite{ritt1950differential}. 
A unique representation of the input-output relation of the system can be obtained by normalising its coefficients so as to yield a monic polynomial. The resulting functions constitute the \textit{exhaustive summary} of the model \cite{walter1982global}. A vector $c(\theta)$ is an exhaustive summary of a model if it only contains the information about $\theta$ that can be inferred from $u(t)$ and $y(t)$. Checking the injectivity of the map $c(\theta)$ amounts to to assessing the identifiability of the model. Several methods can be used for this task. 

\subsection{Overview of tools for analysing structural identifiability and observability \textit{a priori}}\label{sec:tools}

\begin{figure}[ht]
\centering
	\includegraphics[width=0.9\linewidth]{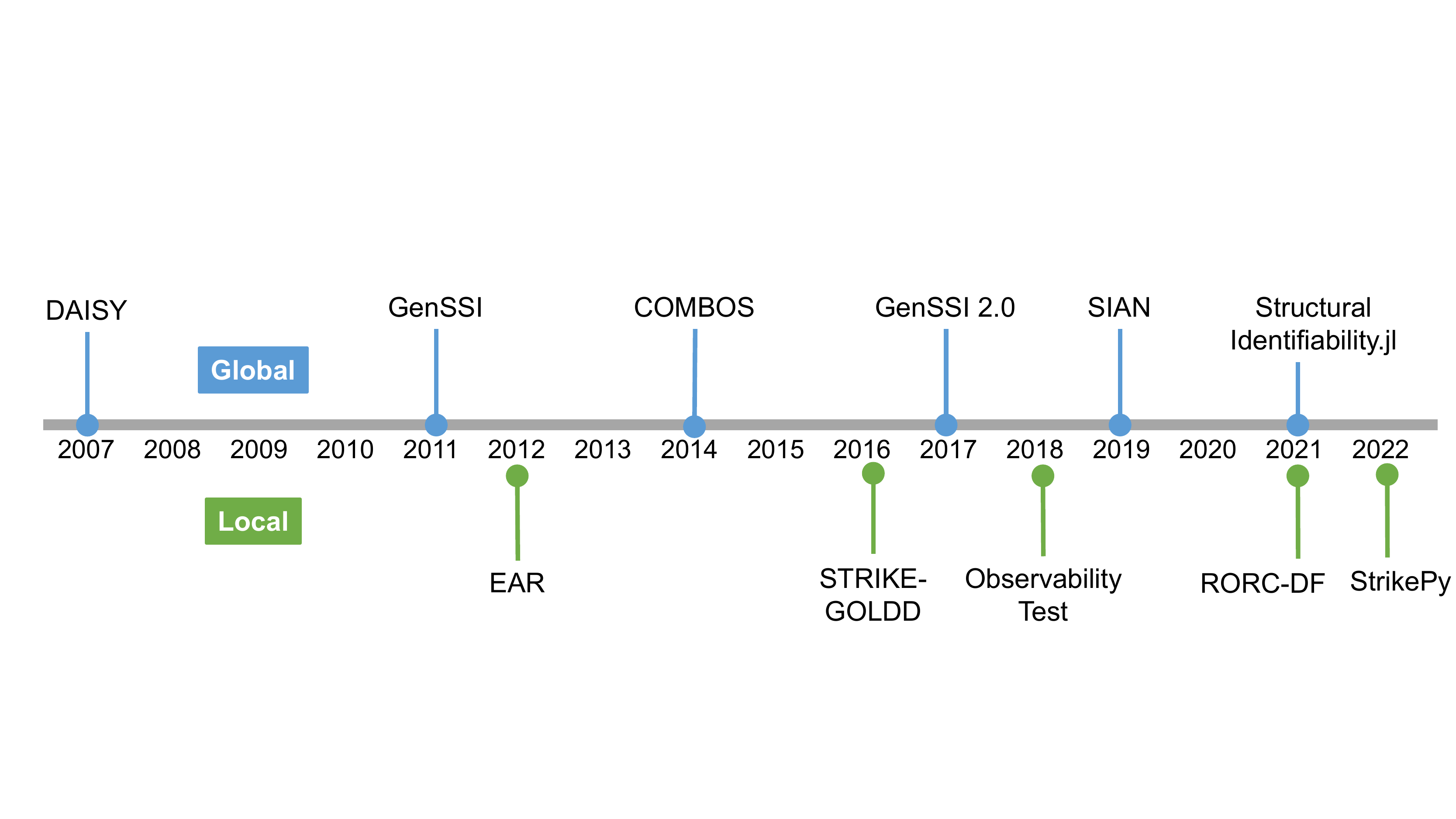}
	\caption{Release dates of the software toolboxes analysed in this paper. Structural global identifiability tools are displayed over the time line, structural local identifiability tools under the time line. For one tool (GenSSI) the years of two main releases are shown, due to the substantial differences between them. For the remaining tools, only the first public release is considered.}
	\label{fig:timeline}
\end{figure}

Despite the significant progress that has taken place in the last decade, structural identifiability analysis is still a challenging subject in systems biology. There are several theoretical approaches and a growing number of tools, none of which is suited for the whole range of nonlinear ODE models developed in this area. 
In this section, we will outline those symbolic methods that have publicly available software implementations. 
They are listed in table \ref{tab:softwares}, and their release timeline is shown in Figure \ref{fig:timeline}.

\begin{table*}[ht]
  \caption{Software tools evaluated in this work. All tools are in principle capable of testing for local identifiability. The ``Features'' columns indicate which methods are capable of the following tasks: analysing global identifiability (``Global''), finding the Lie symmetries in the model equations (``Symmetries''), testing for specific initial conditions (``ICs''), considering models with unknown inputs (``Unknown in''), finding identifiable model reparameterizations (``Reparamet''), analysing non-rational models (``Nonrational''), finding identifiable parameter combinations (``Combin''), and calculating the number of solutions (``$\#$ solutions'').\label{tab:softwares}}
{\begin{tabular}{@{}  p{3.0cm}  p{0.5cm}  p{4.2cm} p{2cm} p{0.2cm} p{0.2cm} p{0.2cm} p{0.2cm} p{0.2cm} p{0.3cm} p{0.2cm} p{0.2cm}  @{}} 
    \toprule
    & & & & \multicolumn{8}{c}{\textbf{Features}} \\
    \textbf{Tool} & \textbf{Ref.} & {\textbf{Web}} & \textbf{Language} & \rot{Global} & {\rot{Symmetries}}  &  \rot{ICs} &  \rot{Unknown in}  &  \rot{Reparamet}  &  \rot{Nonrational}  &  \rot{Combin}  &  \rot{$\#$ solutions}  \\
    \midrule
    ObservabilityTest &   \cite{sedoglavic2002probabilistic} & \footnotesize{\url{https://github.com/sedoglavic/ObservabilityTest}} & Maple &  & \OK & & & \OK & & & \\
    EAR & \cite{karlsson2012efficient} & \footnotesize{\url{http://www.fcc.chalmers.se/software/other-software/identifiabilityanalysis/}} & Mathematica &  & \OK & \OK & & \OK & & & \\    
    STRIKE-GOLDD (FISPO) & \cite{villaverde2016structural}  & \footnotesize{\url{https://github.com/afvillaverde/strike-goldd}} & Matlab &  & \OK & & \OK & \OK & \OK & & \\
    STRIKE-GOLDD (ProbObsTest) & \cite{diaz2022strike}  & \footnotesize{\url{https://github.com/afvillaverde/strike-goldd}} & Matlab &  & \OK & & \OK & \OK & & & \\
    StrikePy & \cite{rey2022strikepy} & \footnotesize{\url{https://pypi.org/project/StrikePy/}} & Python &  &  & & \OK & & & &  \\
    \revision{RORC-DF} & \cite{shi2022efficient} & \footnotesize{\url{https://eng.ox.ac.uk/non-lineardynamics/resources}} & Matlab &  & & & \OK & & & &  \\
    \revision{GenSSI2} & \cite{ligon2018genssi}  & \footnotesize{\url{https://github.com/genssi-developer/GenSSI}} & Matlab & \OK & & \OK & & & \OK & & \\
    SIAN \revision{v1.5} (Maple) & \cite{hong20199sian} & \footnotesize{\url{https://github.com/pogudingleb/SIAN}} & Maple & \OK  & & & & & & &  \OK  \\
    SIAN \revision{v1.1.1} (Julia) & \cite{hong20199sian} & \footnotesize{\url{https://github.com/pogudingleb/SIAN}} & Julia & \OK  & & & & & & &  \OK  \\
    DAISY & \cite{bellu2007daisy} & \footnotesize{\url{https://daisy.dei.unipd.it/}}  & Reduce & \OK &  & \OK & & & & &  \OK  \\
    COMBOS & \cite{meshkat2014finding} & \footnotesize{\url{http://biocyb1.cs.ucla.edu/combos/}} & Maxima (web app) & \OK &  & \OK & & & & \OK &  \OK  \\
    Structural-Identifiability \revision{v0.3.0} &  \cite{dong2022differential}     & \footnotesize{\url{https://github.com/SciML/StructuralIdentifiability.jl}} & Julia & \OK &  & & & & & \OK &  \\
    \bottomrule
    \end{tabular}}{}
\end{table*}

The tools considered in this work can be classified in two broad classes, depending on their approach (differential geometry or differential algebra), although some of them have elements of both -- for example, the generating series approach implemented in GenSSI.
Furthermore, other conceptual differences must also be taken into account, since not all tools provide the same features. For example it was already mentioned that some methods -- in fact, most of them -- are only applicable to rational models. Likewise, some algorithms allow the definition of specific initial conditions, while others do not. Another difference is the possibility of considering models with unknown inputs. Finally, some software tools go beyond structural identifiability and observability analysis, informing about the existence of symmetries, identifiable parameter combinations, or model reparameterizations. 

In the remainder of this section we describe these tools briefly from a conceptual viewpoint. Then, we evaluate their computational performance in Section \ref{sec:results}. 

\subsubsection{DAISY}

DAISY (Differential Algebra for Identifiability of SYstems) was the first symbolic computation tool presented for the analysis of structural global identifiability \cite{bellu2007daisy}. It is a differential algebra software written in REDUCE version 3.8, a free symbolic language. Its algorithm \cite{saccomani2001new} is based on the concept of characteristic set of the ideal generated by the polynomials defining the model. The main idea is to write the input-output relation of the system in implicit form, i.e. as a set of $m$ polynomial differential equations in the variables $(y,u)$, eliminating the dependence on $x$. After ranking the model variables, the characteristic set of the differential ideal is computed with Ritt's pseudodivision algorithm \cite{ritt1950differential}. This yields differential equations whose coefficients depend on the parameter vector $\theta$. To analyse the structural identifiability of the model it is necessary to normalise each of the equations, making it monic. This family of new functions is defined as the exhaustive summary $c(\theta)$, which encapsulates the parameter dependence of the output and whose injectivity $c(\theta)$ has to be checked. In DAISY, the system of algebraic nonlinear equations $c(\theta)$ is solved by the Buchberger algorithm \cite{buchberger1998grobner}. This calculates the Gröbner basis of the system, which provides the number of solutions for each parameter.

\subsubsection{COMBOS}

COMBOS is a web-based application \cite{meshkat2014finding} for structural global identifiability analysis that uses the computer algebra system Maxima. It presents two main developments with respect to DAISY. On the one hand, it provides an installation-free way of analysing structural identifiability. On the other hand, it goes beyond the capabilities of DAISY \cite{meshkat2009algorithm,meshkat2011finding} by providing as additional information the simplest globally identifiable combinations of the unidentifiable parameters. For locally identifiable parameters, COMBOS determines the maximum number of local solutions. 

DAISY and COMBOS differ in the way in which they handle initial conditions. A model that is in principle identifiable for generic initial conditions might be unidentifiable for certain initial conditions from which it is not accessible \cite{saccomani2003parameter}. If we provide specific initial conditions, the results of both software tools are consistent if the system is accessible from those initial conditions, but they may differ in case of inaccessibility. This is because COMBOS, unlike DAISY, does not consider all possible inaccessible cases.

\subsubsection{SIAN}

SIAN (Structural Identifiability ANalyser) is an open-source software tool for structural global identifiability analysis. It combines differential algebra methods with the Taylor series approach \cite{hong2020global,hong20199sian}. SIAN creates a map that binds the parameter values and initial conditions to output functions. By replacing the latter with truncations of their Taylor series, the map is reduced to another map between finite-dimensional spaces. To this end, SIAN determines the order of truncation that contains enough information for the identifiability analysis. The result is correct with a given probability, which is estimated within the algorithm. 

SIAN is available in three implementations: as Maple code, as Julia code, and as a web app in the Maple Cloud server.
In the present study we have focused on assessing the performance of the Maple and Julia implementations, after some preliminary tests showed that the web application is less efficient. 

\subsubsection{StructuralIdentifiability}

StructuralIdentifiability \cite{dong2022differential} is the most recent tool for analysing structural global identifiability. 
While if follows a similar approach as the tools described above, it includes advances with respect to DAISY and COMBOS in the computation of the injectivity test, which is performed in a probabilistic way that increases its efficiency. StructuralIdentifiability.jl is a package implemented in the Julia language as a part of SciML ecosystem, an open source software for scientific machine learning. 

\subsubsection{GenSSI}

GenSSI (Generating Series for testing Structural Identifiability) is a software toolbox for structural global identifiability analysis implemented in Matlab.
It was originally presented in \cite{chics2011genssi}, and a substantially new implementation (GenSSI 2.0) appeared in \cite{ligon2018genssi}.
Its algorithm combines the generation series approach with the so-called identifiability tableaus. 
The generating series approach resembles the power series expansion \cite{pohjanpalo1978system}, which is based on the idea that the Taylor series expansions of the output functions include all the information that is relevant for analysing identifiability. 
Instead of expanding the output functions by Taylor series, GenSSI computes symbolically the successive Lie derivatives of the output functions with respect to parameters and states \cite{walter1982global}. In this way an exhaustive summary is obtained, and from its injectivity the parameter identifiability can be established. In addition, GenSSI provides identifiability tableaus \cite{balsa2010iterative} as a means of determining the number of solutions visually, which is helpful for classifying a parameter as SLI or SGI.

\subsubsection{ObservabilityTest}

ObservabilityTest is a Maple tool for analysing structural local identifiability of rational models. It implements the probabilistic algorithm presented in 2002 by Sedoglavic \cite{sedoglavic2002probabilistic}, which aims at evaluating the observability rank criterion (ORC) efficiently, i.e. in bounded polynomial time. To this end, it avoids the need to compute the Lie derivatives symbolically when building the observability matrix, calculating instead the first terms of a power series expansion and specialising the variables on random integers. Furthermore, when the model is unobservable the power series approach searches for the Lie symmetries that cause the unobservability. 

Since this algorithm is the fastest way of assessing structural local identifiability in rational models, it has been implemented not only in the author's ObservabilityTest (Maple) but also in the aforementioned StructuralIdentifiability.jl (Julia), in EAR (Mathematica), and STRIKE-GOLDD (Matlab). 

\subsubsection{EAR}

The Exact Arithmetic Rank (EAR), also known as IdentifiabilityAnalysis, is a Mathematica application for structural local identifiability analysis \cite{karlsson2012efficient}. At its core lies an implementation of the probabilistic semi-numerical algorithm introduced by Sedoglavic in \cite{sedoglavic2002probabilistic} (which was later implemented in the Maple tool ObservabilityTest). 

As an enhancement over ObservabilityTest, EAR can consider either generic initial conditions (using the ``observability analysis'' function, which treats parameters as constant states) or initial conditions specialised to some numerical value (using the ``identifiability analysis'' function). 
Furthermore, EAR provides functions for finding certain Lie point symmetries in the model, and to compute the minimal output sets for achieving identifiability \cite{anguelova2012minimal}.

\subsubsection{STRIKE-GOLDD}

STRIKE-GOLDD \cite{villaverde2016structural}(STRuctural Identifiability taKen as Extended-Generalized Observability with Lie Derivatives and Decomposition) is a Matlab toolbox for analysing structural local identifiability analysis with the differential geometry approach \cite{villaverde2016structural}.
It provides features such as the search for Lie symmetries and for identifiable reparameterizations \cite{massonis2021autorepar}, or the decomposition of large models to make them tractable.

STRIKE-GOLDD implements three algorithms: 
(1) \underline{FISPO}, which is the most generally applicable, being the only one that can analyse both non-rational models and models with unknown inputs \cite{villaverde2019full}.
(2) \underline{ProbObsTest} \cite{diaz2022strike}, which implements Sedoglavic's algorithm for analysing rational models. With respect to the other implementations in Maple and Mathematica, it presents two developments that enlarge the class of models that it can analyse: on the one hand, it can analyse models with unknown inputs \revision{(similarly to RORC-DF)}; on the other hand, it can automatically transform models with logarithmic, trigonometric, and exponential functions into rational models.
(3) \underline{ORC-DF}, which was originally developed by \cite{maes2019observability}.
In this study we have evaluated the first two algorithms; we refer to these tools as STRIKE-GOLDD (FISPO) and STRIKE-GOLDD (ProbObsTest), respectively. Since our preliminary tests showed that the ORC-DF implementation in STRIKE-GOLDD is less efficient than the \revision{implementation by \cite{maes2019observability}, and this one in turn is less efficient than RORC-DF}, we only considered the latter. 

\subsubsection{StrikePy}

StrikePy is a Python toolbox that analyses structural local identifiability \cite{rey2022strikepy}. It implements the FISPO algorithm of STRIKE-GOLD, but it does not include other features present in that toolbox and is computationally less efficient than the Matlab implementation. However, we evaluate it in this paper since at the moment of writing this article it appears to be the only Python tool for analysing structural identifiability.

\subsubsection{RORC-DF}

\revision{RORC-DF and the previously presented ORC-DF method (Observability Rank Criterion for systems with Direct Feedthrough) adopt a similar approach, but have different applicability. 

ORC-DF \cite{maes2019observability}, which has its own Matlab implementation, can analyse analytical models that are affine in the known and unknown inputs. The term \textit{direct feedthrough} means that the outputs may be functions of both the measured (known) and unmeasured (unknown) inputs. ORC-DF considers the unmeasured inputs and their time derivatives as additional states. 

RORC-DF (rational ORC-DF) \cite{shi2022efficient} was the first extension of Sedoglavic's algorithm to systems with unknown inputs. Unlike ORC-DF, RORC-DF does not require the system to be affine in the inputs, but it introduces the assumption of rational non-linearities. 
In RORC-DF the observability matrix is composed by the coefficients of the power series expansion of the output functions, obtained with Newton's iteration. These computations are carried out using random numerical realizations of the symbolic variables, and applying modular operations to reduce the computational burden. For these reasons RORC-DF is computationally more efficient than ORC-DF. 
}

\section{Results and discussion}\label{sec:results}

We have benchmarked the tools described in Section \ref{sec:tools} by using them to analyse a large and diverse collection of problems from the systems biology literature and related areas. Our benchmark collection is made up of a total of 25 problems created from 21 basic models, which are listed in Table \ref{tab:modelos} along with their references and dimensions (numbers of states, parameters, outputs, and inputs). The collection includes rational and non-rational models, as well as models with and without inputs. For some of the latter we consider both the known and the unknown input case. In regard to their dimensions, the smallest models that we consider have a few parameters and states, while the largest have tens of them. While larger models with hundreds or even thousands of parameters are increasingly common in systems biology, currently existing tools are not capable of analysing them.
In our assessment we consider several criteria, which are discussed in the following subsections. Tables \ref{tab:algdif} and \ref{tab:geomdif} summarise the results of our analyses.

\renewcommand{\arraystretch}{1.3}

\begin{table}[ht]
  \centering
  \caption{List of benchmark models and their main features. The columns display a short name for the model, its original publication, the number of its states, parameters (``param.''), known inputs (``Kn-in''), unknown inputs (``Unk-in''), measured outputs, and whether it is rational or not.}
    \scalebox{1}[1]{
    \begin{tabular}{lccccccc}
    \hline
    \textbf{Short name} & \textbf{Ref.}& \textbf{States} & \textbf{Param.} & \textbf{Kn-in} & \textbf{Unk-in} & \textbf{Outputs} & Rational \\
    \hline
    C2M a & \cite{villaverde2018input} & 2 & 4 & 1 & & 1 & \OK  \\
    C2M b & \cite{villaverde2018input} & 2 & 4 & & & 1  & \OK \\ 
    C2M c & \cite{villaverde2018input} & 2 & 4 &  & 1 & 1  & \OK \\
    Competition & \cite{coleman1972application} & 2 & 6 &  &  & 1  &  \\
    HIV 1 a & \cite{perelson1999mathematical} & 3 & 5 & 1 & & 2  & \OK\\
    HIV 1 b& \cite{perelson1999mathematical} & 3 & 5 & & 1 & 2  & \OK\\
    HIV 2& \cite{perelson1999mathematical} & 4 & 10 &  & & 2  & \OK \\
    HIV 3 & \cite{wodarz1999specific}& 5 & 10 &  & &2  & \OK\\
    NFkB 1& \cite{lipniacki2004mathematical} & 15 & 29 &  & &6  & \OK\\
    NFkB 2& \cite{lipniacki2004mathematical} & 15 & 6 & 1 & & 6 & \OK \\
    Phosphorylation& \cite{conradi2018dynamics} & 6 & 6 &  && 2  & \OK \\
    PK 1& \cite{raksanyi1986utilisation} & 4 & 9 &  && 2 & \OK \\
    PK 2& \cite{verdiere2005identifiability} & 4 & 9 &  & &1  & \OK \\
    Ruminal lipolysis& \cite{moate2008kinetics} & 5 & 4 &  & & 3   & \OK \\
    Tumor& \cite{thomas1989effect} & 5 & 5 &  & &1 & \OK \\
    MAPK & \cite{nguyen2015dyvipac} & 3 & 14 &  & &3 &  \\
    \textit{A. thaliana}& \cite{locke2005modelling} & 7 & 29 & 1 && 2 &  \\
    Toggle switch a & \cite{lugagne2017balancing} & 2 & 10 & 2 & & 2  &   \\
    Toggle switch b & \cite{lugagne2017balancing} & 2 & 10 & & 2 & 2  &   \\ 
    JAK-STAT 1 & \cite{raia2011dynamic}& 10 & 23 & 1 & &8 & \OK \\
    JAK-STAT 2 & \cite{bachmann2011division}& 25 & 24 &  && 14  & \OK\\   
    $\beta$IG& \cite{topp2000model} & 3 & 5 & \textcolor[rgb]{ 1,  0,  0}{}1 && 1 & \OK \\
    SIRS with forcing& \cite{weber2001modeling} & 5 & 13 & 1 & & 2 & \OK \\
    Cholera & \cite{lee2017model} & 4 & 7 &  & &2 & \OK \\
    Gene p53& \cite{distefano2015dynamic} & 4 & 25 & 1 & & 4 & \OK \\
    \hline
    \end{tabular}}%
  \label{tab:modelos}%
\end{table}%

\begin{table}[htbp]
  \centering
  \caption{Structural Global Identifiability tools: summary of results. The table entries display the runtimes for each benchmark model, in seconds. An asterisk (*) next to a value denotes that the result is thought to be wrong, while a diamond ($^{\diamond}$) denotes that the correctness of the result is unclear. \revision{In one case (the JAK-STAT 2 model) no tool was capable of assessing global identifiability, but two of them managed to provide at least local results; they are indicated with a subscript ($_L$).}}
  \scalebox{1}[1]{
    \begin{tabular}{lcccccc}
    \hline
    &  \textbf{DAISY} & \textbf{GenSSI} & \multicolumn{1}{p{1.5cm}}{\textbf{SIAN (Maple)}}  & \multicolumn{1}{p{1.5cm}}{\textbf{SIAN (Julia)}} & \textbf{COMBOS} & \multicolumn{1}{p{2.5cm}}{\textbf{Structural Identifiability}} \\
    \hline
    C2M a &  0.34 &  2.81 &  0.358 & 10.27 & 0.31  & 33.91\\
    C2M b & 0.24* & 7.08  & 0.28  & 10.27 & 0.98  & 33.71\\
    C2M c & N/A & N/A & N/A & N/A & N/A  & N/A\\
    Competition & N/A & Error & N/A & N/A & N/A  & N/A\\
    HIV 1 a & 0.13* & 0.73 & 0.80 & 12.09 & 0.83 & 32.02\\
    HIV 1 b & N/A & N/A & N/A & N/A & N/A  & N/A\\
    HIV 2 & 0.448*  & 966.66* & 6.687 & 10.63 & 36.23 & 31.60\\
    HIV 3 & 6.66 & 751.74* & 31.34 & 14.32 & Error & 32.78\\
    NFkB 1 & Error & 6722.98* & 3867.02 & Error & Error & Error\\
    NFkB 2  & Error & 660.36 & 3690.70 & 244.27 & Error & Error\\
    Phosphorilation & 19.43 & 974.61 & 5.23 & 13.64 & Error & 35.02\\
    PK 1 & 0.31 & 14.58* & 1.48 & 12.26 & 5.72* & 34.39\\
    PK 2  & Error & 5082.68* & Error & Error & Error  & 84.86\\
    Ruminal lipolysis & 0.12 & 1.46 & 0.95 & 14.12 & 220.00 & 34.86\\
    Tumor & 8.91 & 1433.22 & 940.55 & 404.96 & 6735.38  & 34.87\\
    MAPK & 0.15* & 27.80 & N/A & N/A & Error & N/A\\
    \textit{A. thaliana} & N/A & $6356.60^{\diamond}$  & N/A & N/A & N/A & N/A\\
    Toggle switch a & N/A & N/A & N/A & N/A & N/A & N/A\\
    Toggle switch b & N/A & N/A & N/A & N/A & N/A & N/A\\
    JAK-STAT 1 & Error & 23284.00* & 246.48 & 40.90 &  Error & 62.55\\
    JAK-STAT 2 & Error & N/A & $115200.00^{\diamond}_{\revision{L}}$ & Error &  Error & $37.21^{\diamond}_{\revision{L}}$ \\
    $\beta$IG  & 0.09* & 16999.00* & 6.54 & 11.33 & Error & 31.65\\
    SIRS with forcing & Error & 648.26* & 10.94 & 12.13 &  Error & 41.56\\
    Cholera & Error & 361.08 & 380.67 & 56.52 & Error & 34.72\\
    Gene p53 & 0.792 & Error & 221.73 & 9331.52 & 1339.38 & 33.15\\
    \hline
    \end{tabular}}%
  \label{tab:algdif}%
\end{table}%

\begin{table}[htbp]
  \centering
  \caption{Structural Local Identifiability tools: summary of results. The table entries display the runtimes for each benchmark model, in seconds. An asterisk (*) next to a value denotes that the result is thought to be wrong, while a diamond ($^{\diamond}$) denotes that the correctness of the result is unclear.}
  \scalebox{1}[1]{
    \begin{tabular}{lcccccc}
     \hline
    & \multicolumn{1}{p{2cm}}{\textbf{STRIKE-GOLDD (FISPO)}} & \multicolumn{1}{p{2.7cm}}{\textbf{STRIKE-GOLDD (ProbObsTest)}} & \textbf{StrikePy} & \multicolumn{1}{p{2cm}}{\textbf{Observability Test}} & \textbf{\revision{RORC-DF}} & \textbf{EAR} \\
    \hline
    C2M a                & 0.63  & 1.58 & 2.77 & 0.31*e-1 &  3.40 & 0.05 \\
    C2M b                & 1.17  & 1.72 & 7.97 & 0.46*e-1  & 4.58 & 0.11 \\
    C2M c                & 12.55 & 4.30 & 37.90 & N/A & 16.21  & N/A\\
    Competition          & 1696.29 & 7.42* & Error & N/A & N/A  & N/A\\
    HIV 1 a              & 0.74  & 3.96 & 0.52 & 0.09 & \revision{7.74} & 0.19 \\
    HIV 1 b              & \revision{$2.23^{\diamond}$}  & \revision{$8.65^{\diamond}$} & \revision{$24.31^{\diamond}$} & N/A & \revision{$11.27^{\diamond}$}  & N/A\\
    HIV 2                & 29.79 & 10.34 & 1685.76 & 0.22  & \revision{40.59} & 0.6095 \\
    HIV 3                & 8528.00 & 12.76 & Error & 0.20 & 36.42 & 1.25 \\
    NFkB 1               & 33345.00 & 304.40 & Error & 8.42 &  \revision{11666.91} & 24.37 \\
    NFkB 2               & 1007.00 & 329.83 & Error & 3.14 & \revision{1138.97} & 6.26 \\
    Phosphorilation      & 1.87    & 13.41 & 32.40 & 0.16 & 28.05 & 0.91 \\
    PK 1                 & 2.69    & 6.41 & 198.96 & 0.14 & 34.00 & 0.58 \\
    PK 2                 & Error   & 16.41 & Error & 0.14 & 14.87 & 0.58\\
    Ruminal lipolysis    & 0.74 & 17.07 & 22.83 & 0.13 & \revision{12.95} & 0.38 \\
    Tumor                & 24.86 & 8.66 & 636.55 & 0.17 & 140.13 & 1.00 \\
    MAPK                 & 94.219 & N/A & Error & N/A  & N/A & N/A \\
    \textit{A. thaliana} & $167769.33^{\diamond}$ & N/A & N/A & N/A  & N/A & N/A \\
    Toggle switch a      & 62.85 &  N/A & Error & N/A & N/A & N/A \\
    Toggle switch b      & 29.497 &  N/A & Error & N/A & N/A & N/A \\
    JAK-STAT 1           & 31.26 & 203.26*  & Error & 2.00  & \revision{1723.97} & 4.92* \\
    JAK-STAT 2           & $146450.00^{\diamond}$ & $2318.46^{\diamond}$  & Error & $35.74^{\diamond}$  & \revision{$86333.30^{\diamond}$} & Error \\
    $\beta$IG            & 2059.89 & 16.10  & Error & 0.08 & 9.60 & 0.7663 \\
    SIRS with forcing    & 87.98 & 9.20  & 2836.98 & 0.13  & \revision{31.05} & 0.54 \\
    Cholera              & 162.26 &  8.02 & 210.39 & 0.11 & \revision{14.02} & 0.46 \\
    Gene p53             & 308.05 & 112.53 & 29193.25 & 0.34 & \revision{45.35} & 3.32 \\
    \hline
    \end{tabular}}%
  \label{tab:geomdif}%
\end{table}%

\subsection{Software accessibility and usability} 

\paragraph{Accessibility.}
The first issue with which the prospective user of a software tool needs to deal is obtaining it \cite{lee2021barely}. Most of the toolboxes studied here are directly and freely available on their own website or in GitHub. The two exceptions are EAR and DAISY, which are available upon request by email to their authors.

\paragraph{Usability.}
The second issue is learning how to use the software and how to apply it to one's needs \cite{list2017ten}. \revision{All the toolboxes evaluated here} provide either a README file or a user manual, or both. 
As for debugging, some programming environments such as Matlab, Julia, and Maple provide detailed reports of the problems encountered when executing a code. Other environments, namely Reduce, Mathematica, and the COMBOS WebApp, do not specify the cause of the problem. 
Being able to track the source of an error is particularly important when working with large models, and when the user is not familiar with a program.

\subsection{Possibility of performing a given analysis} 

\paragraph{Allowed model types.}
This criterion considers the theoretical possibility of performing the analysis. 
The most common limitation regards the analysis of non-rational models, which can only be performed by STRIKE-GOLDD (FISPO), StrikePy, and GenSSI. ProbObsTest, DAISY, and COMBOS can deal with rational functions as long as they can be transformed into polynomial functions. 
In the case of functions with non-integer exponents (such as JAK-STAT 1 and $\beta$IG), their analysis with SIAN, StructuralIdentifiability, ObservabilityTest, and EAR requires approximating their values to the closest integer. While in general this change should not alter the identifiability results, it can reduce computation times. Hence, if a model is modified in this way, it should also be modified when analysing it with other methods, in order to ensure a fair comparison. 
Another common limitation concerns models with unknown inputs. Only four methods can lead with this class of models, all of which use local approaches: \revision{RORC-DF}, STRIKE-GOLDD (FISPO and ProbObsTest), and StrikePy.

\paragraph{Allowed analyses.}
Regarding the type of analysis, the most common difference is that some methods can only determine structural local identifiability, while others can also analyse global identifiability.
Furthermore, some tools (StrikePy and \revision{RORC-DF}) provide only identifiability and observability results, while others also search for symmetries, identifiable parameter combinations, or reparameterizations.
Table \ref{tab:softwares} lists the main features of each tool. 
Some global tools such as SIAN, ObservabilityTest, DAISY, COMBOS, and GenSSI provide information about the number of local solutions. Some local tools such as EAR, STRIKE-GOLDD and ObservabilityTest assist in finding symmetries and model reparameterizations. 

\paragraph{Computational feasibility.}
The feasibility of the analysis in practice must also be considered: even if a tool can analyse a given model in principle, it may not be able to do so due to computational limitations.
This is reflected in the number of errors shown in the table \ref{tab:algdif} and \ref{tab:geomdif}. An error in these tables means that the tool was not able to analyse the model due to computational limitations. The most limited tools in this regard are StrikePy (due to the use of Python) and COMBOS (whose limitations may stem from the WebApp server). 
Another problematic tool in this regard is DAISY, which yields less errors than StrikePy and COMBOS but struggles with medium-sized models. %
It should be noted that some of the table entries are labelled as errors because in our tests we limit the calculations to 48 hours, stopping the analysis if it did not finish in that time. Both DAISY and StrikePy were affected by this bound.

\subsection{Results} 

\paragraph{Correctness.} Even when a tool has produced results for a given model, they may not always be correct.
For some models we found discrepancies among the results of several tools. In such cases there was typically a clear consensus among methods, with only one or two methods that disagree with the common solution; in this case we assumed that the consensus solution is the true one, and we marked the wrong solutions with an asterisk (*) in Tables \ref{tab:algdif}, \ref{tab:geomdif}. However, in \revision{three cases (HIV 1 b,} JAK-STAT 2 and \textit{A. thaliana}) there was not a clear majority; in these cases we did not make any assumptions about correctness, and we wrote a diamond  ($^{\diamond}$) next to all results in the tables.
Under these assumptions, we found that \revision{five methods did not produce any wrong result: SIAN, StructuralIdentifiability, ObservabilityTest, RORC-DF, and STRIKE-GOLDD (FISPO)}. Two algorithms, EAR and STRIKE-GOLDD (ProbObsTest), yielded wrong results for JAK-STAT 1. In this regard, we have realised that this result depends on the choice of prime number used by these methods to specialise the variables on random numbers; if we select the same prime number we obtain the same result. 
Additionally, STRIKE-GOLDD (ProbObsTest) yielded a wrong result for the Competition model, which could only be analysed with this method and with STRIKE-GOLDD (FISPO). This case study illustrates the following issue: due to the presence of logarithmic terms, methods such as ProbObsTest must transform the model into polynomial form in order to analyse it; however, the transformed model does not necessarily preserve the properties of the original model. 
\revision{Two} other tools, DAISY, and GenSSI, produced wrong results for a number of case studies. 

\paragraph{Computational performance.} 
Even when two tools agree on the result, their computational costs may be very different. Tables \ref{tab:algdif} and \ref{tab:geomdif} show CPU times, which we have used as the main measure of this criterion. 
They depend on the programming environment and the algorithm. Clearly, the fastest algorithm in our tests was the one developed by Sedoglavic \cite{sedoglavic2002probabilistic}, which is implemented with some variations in \revision{four toolboxes -- ObservabilityTest, EAR, RORC-DF, and STRIKE-GOLDD (ProbObsTest)} -- programmed in three different languages -- Maple, Mathematica, Matlab. The fastest implementation was the Maple one, followed by the Mathematica one.
Those three tools are restricted to structural \textit{local} identifiability analysis (global tools are usually slower). Among the remaining local tools, the next two in terms of computational efficiency were STRIKE-GOLDD (FISPO) and \revision{RORC-DF}
. The slowest tool of all was StrikePy.
Among \textit{global} tools, GenSSI yielded the largest CPU times; DAISY was on average faster than GenSSI, although it managed to complete the analysis of fewer models (probably due to the 48 hour limit that we imposed to the calculations). We found a similar, but even more pronounced effect for COMBOS.
In comparison, SIAN and StructuralIdentifiability performed very well.
The computation times of StructuralIdentifiability were remarkably similar for most models, regardless of their size. 
We tested two implementations of SIAN, in Maple and in Julia. The Maple implementation was faster than the Julia one for smaller models, and slower for larger models. 
Julia uses Just-in-time compilation, where each function is compiled the first time it is called. Therefore, the computation times in table \ref{tab:algdif} count this compilation time together with loading the package, which may be about 20-30 seconds.

\section{Conclusions}

Here we have presented a critical analysis and computational benchmarking of the existing software tools for analysing structural identifiability and observability. Our results have revealed their relative strengths and weaknesses. Below, we outline some guidelines for choosing the most appropriate tool for a given problem, and then we provide specific recommendations. Finally, we conclude with a few words about possible directions for future research.

\subsection{General guidelines}

First, the decision process must consider the \textit{type of model} that has to be analysed, since not all methods can be applied to all models. If it is a rational model without unknown inputs (a common situation in systems biology as well as in other areas), all methods can be applied. However, for other model types the choice of applicable methods is reduced, as can be seen in Table \ref{tab:softwares}. 

Second, the user must decide whether to assess \textit{global (SGI) or local (SLI)} structural identifiability, if both approaches are applicable to the model. SGI implies SLI but the opposite is not true; while it is often the case that a SLI model is indeed SGI, some counter-examples have been reported. The extent to which the distinction between local and global identifiability is relevant in biological applications is worthy of further investigation. If it is not necessary to assess SGI, it may be convenient to resort to SLI methods, since they are usually computationally cheaper than SGI methods.

Another factor is the \textit{software environment}. We have benchmarked tools written in six different programming languages, as well as some web-based applications that do not require the installation of specific software. While the array of available methods is reasonably large, for a given language the number of possible choices is usually restricted to two or at most three, and sometimes only one. Thus, the (in)convenience of reimplementing the model in a different language needs to be taken into account when choosing a software tool. This is especially important if the structural identifiability analysis is performed as part of a larger computational pipeline for model building and exploitation, which is a typical scenario. In this case, it is desirable to be able to perform all analyses within the same software environment.
It should also be taken into account some of the environments are proprietary software (Matlab, Mathematica, Maple), and therefore not available to every user.

Finally, some tools provide \textit{additional features}, which can be used to reformulate a model if it is unidentifiable. Such features include the search for symmetries in the model equations, identifiable parameter combinations, and identifiable model reparameterizations.

After considering the aforementioned factors, there may be several tools that meet the requirements for the problem at hand. In this case, the user may choose the one with the lowest \textit{computational cost}. As our results have shown, computation times can vary greatly from one tool to another.

\subsection{Recommendations}

From the above discussion it is apparent that the choice of the most appropriate tool is strongly problem-dependent. While every tool has its particular merits, not all of them are equally useful. Hence we would like to provide some final recommendations, which can be summarised as follows.

Within the tools that analyse structural \textit{global} identifiability, there is a clear distinction between the more recent ones and the older ones. The oldest one, DAISY, was the first tool of its kind to be made publicly available; however, the array of models that it is capable of analysing is currently smaller than that of other tools. The next one, COMBOS, was a welcome innovation at the time of its release thanks to its web app implementation; however, it exhibits similar or worse limitations as DAISY. On the other hand, the two most recent methods, SIAN and StructuralIdentifiability, do not share the limitations of the oldest ones. GenSSI lies somewhere in the middle of both groups. Therefore, we recommend using either SIAN (Maple) or StructuralIdentifiability (Julia) for analysing structural global identifiability. The choice between them can boil down to a matter of programming language.

The tools that analyse structural \textit{local} identifiability do not exhibit the same differences in performance between older and newer implementations. We can classify them in two groups, depending on whether they use some version of Sedoglavic's algorithm -- ObservabilityTest, EAR, STRIKE-GOLDD (ProbObsTest), and \revision{RORC-DF} -- or not -- STRIKE-GOLDD (FISPO), StrikePy. The first group yields faster calculations than the second one, but it cannot analyse non-rational models. For the analysis of \textit{rational} models we recommend, in order of computational efficiency, (1) ObservabilityTest, which is by far the fastest tool; (2) EAR; (3) STRIKE-GOLDD (ProbObsTest) \revision{or RORC-DF}. Naturally, the final decision depends on the access to Maple, Mathematica, and Matlab environments.
For the analysis of non-rational models, STRIKE-GOLDD (FISPO) is in some cases the only available option. StrikePy does not outperform other tools and, given its limitations, it should be avoided unless it is necessary to perform the analysis in Python. 

\subsection{Directions for future research}

As our results have shown, recent developments have yielded considerable advances in the available tools for structural identifiability analysis. However, further improvements are still needed to facilitate the analysis of more models, as they tend to become larger and more complex. In this regard, a promising line of work would be to implement more features in the Julia programming language, due to its computational efficiency. 
It should also be noted that all the tools considered in this paper analyse ODE models. While they are the most common ones in systems biology, other types of models are also useful, such as those with partial differential equations or stochastic dynamics. The development of tools for their analysis would greatly broaden the applicability of structural identifiability analysis.

\section*{Availability and implementation} Implementations of all the case studies in all of the toolboxes can be download from \url{https://github.com/Xabo-RB/Benchmarking_files}.

\section*{Funding}
This research has received support 
from grant ED431F 2021/003 funded by Conseller\'ia de Cultura, Educaci\'on e Ordenaci\'on Universitaria, Xunta de Galicia; 
from grant PID2020-113992RA-I00 funded by MCIN/AEI/ 10.13039/501100011033 (PREDYCTBIO);
and from grant RYC-2019-027537-I funded by MCIN/AEI/ 10.13039/501100011033 and by ``ESF Investing in your future''.
The funding bodies played no role in the design of the study, the collection and analysis of data, or in the writing of the manuscript.



\begin{thebibliography}{10}
	
	\bibitem{anguelova2004nonlinear}
	M.~Anguelova.
	\newblock {\em Nonlinear observability and identifiability: general theory and
		a case study of a kinetic model for S. cerevisiae}.
	\newblock Chalmers Tekniska Hogskola (Sweden), 2004.
	
	\bibitem{anguelova2012minimal}
	M.~Anguelova, J.~Karlsson, and M.~Jirstrand.
	\newblock Minimal output sets for identifiability.
	\newblock {\em Mathematical biosciences}, 239(1):139--153, 2012.
	
	\bibitem{anstett2020priori}
	F.~Anstett-Collin, L.~Denis-Vidal, and G.~Mill{\'e}rioux.
	\newblock A priori identifiability: An overview on definitions and approaches.
	\newblock {\em Annual Reviews in Control}, 50:139--149, 2020.
	
	\bibitem{apgar2010sloppy}
	J.~F. Apgar, D.~K. Witmer, F.~M. White, and B.~Tidor.
	\newblock Sloppy models, parameter uncertainty, and the role of experimental
	design.
	\newblock {\em Molecular BioSystems}, 6(10):1890--1900, 2010.
	
	\bibitem{bachmann2011division}
	J.~Bachmann, A.~Raue, M.~Schilling, M.~E. B{\"o}hm, C.~Kreutz, D.~Kaschek,
	H.~Busch, N.~Gretz, W.~D. Lehmann, J.~Timmer, et~al.
	\newblock Division of labor by dual feedback regulators controls {JAK2/STAT5}
	signaling over broad ligand range.
	\newblock {\em Molecular systems biology}, 7(1):516, 2011.
	
	\bibitem{balsa2010iterative}
	E.~Balsa-Canto, A.~A. Alonso, and J.~R. Banga.
	\newblock An iterative identification procedure for dynamic modeling of
	biochemical networks.
	\newblock {\em BMC systems biology}, 4(1):1--18, 2010.
	
	\bibitem{bellman1970structural}
	R.~Bellman and K.~J. {\AA}str{\"o}m.
	\newblock On structural identifiability.
	\newblock {\em Mathematical biosciences}, 7(3-4):329--339, 1970.
	
	\bibitem{bellu2007daisy}
	G.~Bellu, M.~P. Saccomani, S.~Audoly, and L.~D’Angi{\`o}.
	\newblock {DAISY}: A new software tool to test global identifiability of
	biological and physiological systems.
	\newblock {\em Computer methods and programs in biomedicine}, 88(1):52--61,
	2007.
	
	\bibitem{buchberger1998grobner}
	B.~Buchberger and F.~Winkler.
	\newblock {\em Gr{\"o}bner bases and applications}, volume~17.
	\newblock Cambridge University Press Cambridge, 1998.
	
	\bibitem{chics2011genssi}
	O.~Chi{\c{s}}, J.~R. Banga, and E.~Balsa-Canto.
	\newblock {GenSSI}: a software toolbox for structural identifiability analysis
	of biological models.
	\newblock {\em Bioinformatics}, 27(18):2610--2611, 2011.
	
	\bibitem{chis2011structural}
	O.-T. Chis, J.~R. Banga, and E.~Balsa-Canto.
	\newblock Structural identifiability of systems biology models: a critical
	comparison of methods.
	\newblock {\em PloS one}, 6(11):e27755, 2011.
	
	\bibitem{coleman1972application}
	T.~Coleman and J.~Gomatam.
	\newblock Application of a new model of species competition to {Drosophila}.
	\newblock {\em Nature New Biology}, 239(95):251--253, 1972.
	
	\bibitem{conradi2018dynamics}
	C.~Conradi and A.~Shiu.
	\newblock Dynamics of posttranslational modification systems: Recent progress
	and future directions.
	\newblock {\em Biophysical journal}, 114(3):507--515, 2018.
	
	\bibitem{diaz2022strike}
	S.~D\'iaz, X.~Rey, and A.~F. Villaverde.
	\newblock {STRIKE-GOLDD} 4.0: user-friendly, efficient analysis of structural
	identifiability and observability.
	\newblock {\em arXiv preprint arXiv:2207.07346}, 2022.
	
	\bibitem{distefano2015dynamic}
	J.~Distefano.
	\newblock {\em Dynamic systems biology modeling and simulation}.
	\newblock Academic Press, 2015.
	
	\bibitem{dong2022differential}
	R.~Dong, C.~Goodbrake, H.~A. Harrington, and G.~Pogudin.
	\newblock Differential elimination for dynamical models via projections with
	applications to structural identifiability.
	\newblock {\em arXiv preprint arXiv:2111.00991}, 2022.
	
	\bibitem{eisenberg2017confidence}
	M.~C. Eisenberg and H.~V. Jain.
	\newblock A confidence building exercise in data and identifiability: Modeling
	cancer chemotherapy as a case study.
	\newblock {\em Journal of theoretical biology}, 431:63--78, 2017.
	
	\bibitem{hermann1977nonlinear}
	R.~Hermann and A.~Krener.
	\newblock Nonlinear controllability and observability.
	\newblock {\em IEEE Transactions on automatic control}, 22(5):728--740, 1977.
	
	\bibitem{hong20199sian}
	H.~Hong, A.~Ovchinnikov, G.~Pogudin, and C.~Yap.
	\newblock {SIAN}: software for structural identifiability analysis of ode
	models.
	\newblock {\em Bioinformatics}, 35(16):2873--2874, 2019.
	
	\bibitem{hong2020global}
	H.~Hong, A.~Ovchinnikov, G.~Pogudin, and C.~Yap.
	\newblock Global identifiability of differential models.
	\newblock {\em Communications on Pure and Applied Mathematics},
	73(9):1831--1879, 2020.
	
	\bibitem{janzen2016parameter}
	D.~L. Janz{\'e}n, L.~Bergenholm, M.~Jirstrand, J.~Parkinson, J.~Yates, N.~D.
	Evans, and M.~J. Chappell.
	\newblock Parameter identifiability of fundamental pharmacodynamic models.
	\newblock {\em Frontiers in physiology}, 7:590, 2016.
	
	\bibitem{karlsson2012efficient}
	J.~Karlsson, M.~Anguelova, and M.~Jirstrand.
	\newblock An efficient method for structural identifiability analysis of large
	dynamic systems.
	\newblock {\em IFAC proceedings volumes}, 45(16):941--946, 2012.
	
	\bibitem{lee2017model}
	E.~C. Lee, M.~R. Kelly~Jr, B.~M. Ochocki, S.~M. Akinwumi, K.~E. Hamre, J.~H.
	Tien, and M.~C. Eisenberg.
	\newblock Model distinguishability and inference robustness in mechanisms of
	cholera transmission and loss of immunity.
	\newblock {\em Journal of theoretical biology}, 420:68--81, 2017.
	
	\bibitem{lee2021barely}
	G.~Lee, S.~Bacon, I.~Bush, L.~Fortunato, D.~Gavaghan, T.~Lestang, C.~Morton,
	M.~Robinson, P.~Rocca-Serra, S.-A. Sansone, et~al.
	\newblock Barely sufficient practices in scientific computing.
	\newblock {\em Patterns}, 2(2):100206, 2021.
	
	\bibitem{ligon2018genssi}
	T.~S. Ligon, F.~Fr{\"o}hlich, O.~T. Chi{\c{s}}, J.~R. Banga, E.~Balsa-Canto,
	and J.~Hasenauer.
	\newblock {GenSSI} 2.0: multi-experiment structural identifiability analysis of
	sbml models.
	\newblock {\em Bioinformatics}, 34(8):1421--1423, 2018.
	
	\bibitem{lipniacki2004mathematical}
	T.~Lipniacki, P.~Paszek, A.~R. Brasier, B.~Luxon, and M.~Kimmel.
	\newblock Mathematical model of {NF-$\kappa$B} regulatory module.
	\newblock {\em Journal of theoretical biology}, 228(2):195--215, 2004.
	
	\bibitem{list2017ten}
	M.~List, P.~Ebert, and F.~Albrecht.
	\newblock Ten simple rules for developing usable software in computational
	biology.
	\newblock {\em PLoS computational biology}, 13(1):e1005265, 2017.
	
	\bibitem{ljung1994global}
	L.~Ljung and T.~Glad.
	\newblock On global identifiability for arbitrary model parametrizations.
	\newblock {\em Automatica}, 30(2):265--276, 1994.
	
	\bibitem{locke2005modelling}
	J.~C. Locke, A.~J. Millar, and M.~S. Turner.
	\newblock Modelling genetic networks with noisy and varied experimental data:
	the circadian clock in {Arabidopsis thaliana}.
	\newblock {\em Journal of theoretical biology}, 234(3):383--393, 2005.
	
	\bibitem{lugagne2017balancing}
	J.-B. Lugagne, S.~S. Carrillo, M.~Kirch, A.~K{\"o}hler, G.~Batt, and P.~Hersen.
	\newblock Balancing a genetic toggle switch by real-time feedback control and
	periodic forcing.
	\newblock {\em Nature communications}, 8(1):1--8, 2017.
	
	\bibitem{maes2019observability}
	K.~Maes, M.~Chatzis, and G.~Lombaert.
	\newblock Observability of nonlinear systems with unmeasured inputs.
	\newblock {\em Mechanical Systems and Signal Processing}, 130:378--394, 2019.
	
	\bibitem{massonis2021autorepar}
	G.~Massonis, J.~R. Banga, and A.~F. Villaverde.
	\newblock Autorepar: A method to obtain identifiable and observable
	reparameterizations of dynamic models with mechanistic insights.
	\newblock {\em International Journal of Robust and Nonlinear Control}, 2021.
	
	\bibitem{merkt2015higher}
	B.~Merkt, J.~Timmer, and D.~Kaschek.
	\newblock Higher-order {Lie} symmetries in identifiability and predictability
	analysis of dynamic models.
	\newblock {\em Physical Review E}, 92(1):012920, 2015.
	
	\bibitem{meshkat2011finding}
	N.~Meshkat, C.~Anderson, and J.~J. DiStefano~III.
	\newblock Finding identifiable parameter combinations in nonlinear {ODE} models
	and the rational reparameterization of their input--output equations.
	\newblock {\em Mathematical biosciences}, 233(1):19--31, 2011.
	
	\bibitem{meshkat2009algorithm}
	N.~Meshkat, M.~Eisenberg, and J.~J. DiStefano~III.
	\newblock An algorithm for finding globally identifiable parameter combinations
	of nonlinear {ODE} models using {Gr{\"o}bner Bases}.
	\newblock {\em Mathematical biosciences}, 222(2):61--72, 2009.
	
	\bibitem{meshkat2014finding}
	N.~Meshkat, C.~E.-z. Kuo, and J.~DiStefano~III.
	\newblock On finding and using identifiable parameter combinations in nonlinear
	dynamic systems biology models and {COMBOS}: a novel web implementation.
	\newblock {\em PLoS One}, 9(10):e110261, 2014.
	
	\bibitem{miao2011identifiability}
	H.~Miao, X.~Xia, A.~S. Perelson, and H.~Wu.
	\newblock On identifiability of nonlinear {ODE} models and applications in
	viral dynamics.
	\newblock {\em SIAM review}, 53(1):3--39, 2011.
	
	\bibitem{moate2008kinetics}
	P.~Moate, R.~Boston, T.~Jenkins, and I.~Lean.
	\newblock Kinetics of ruminal lipolysis of triacylglycerol and biohydrogenation
	of long-chain fatty acids: new insights from old data.
	\newblock {\em Journal of Dairy Science}, 91(2):731--742, 2008.
	
	\bibitem{munoz2018or}
	R.~Mu{\~n}oz-Tamayo, L.~Puillet, J.-B. Daniel, D.~Sauvant, O.~Martin,
	M.~Taghipoor, and P.~Blavy.
	\newblock To be or not to be an identifiable model. is this a relevant question
	in animal science modelling?
	\newblock {\em Animal}, 12(4):701--712, 2018.
	
	\bibitem{nguyen2015dyvipac}
	L.~K. Nguyen, A.~Degasperi, P.~Cotter, and B.~N. Kholodenko.
	\newblock Dyvipac: an integrated analysis and visualisation framework to probe
	multi-dimensional biological networks.
	\newblock {\em Scientific reports}, 5(1):1--17, 2015.
	
	\bibitem{perelson1999mathematical}
	A.~S. Perelson and P.~W. Nelson.
	\newblock Mathematical analysis of hiv-1 dynamics in vivo.
	\newblock {\em SIAM review}, 41(1):3--44, 1999.
	
	\bibitem{pohjanpalo1978system}
	H.~Pohjanpalo.
	\newblock System identifiability based on the power series expansion of the
	solution.
	\newblock {\em Mathematical biosciences}, 41(1-2):21--33, 1978.
	
	\bibitem{raia2011dynamic}
	V.~Raia, M.~Schilling, M.~B{\"o}hm, B.~Hahn, A.~Kowarsch, A.~Raue, C.~Sticht,
	S.~Bohl, M.~Saile, P.~M{\"o}ller, et~al.
	\newblock Dynamic mathematical modeling of il13-induced signaling in hodgkin
	and primary mediastinal b-cell lymphoma allows prediction of therapeutic
	targets.
	\newblock {\em Cancer research}, 71(3):693--704, 2011.
	
	\bibitem{raksanyi1986utilisation}
	A.~Raksanyi.
	\newblock {\em Utilisation du calcul formel pour l'{\'e}tude des syst{\`e}mes
		d'{\'e}quations polynomiales (applications en mod{\'e}lisation)}.
	\newblock PhD thesis, Paris 9, 1986.
	
	\bibitem{raue2014comparison}
	A.~Raue, J.~Karlsson, M.~P. Saccomani, M.~Jirstrand, and J.~Timmer.
	\newblock Comparison of approaches for parameter identifiability analysis of
	biological systems.
	\newblock {\em Bioinformatics}, 30(10):1440--1448, 2014.
	
	\bibitem{raue2009structural}
	A.~Raue, C.~Kreutz, T.~Maiwald, J.~Bachmann, M.~Schilling, U.~Klingm{\"u}ller,
	and J.~Timmer.
	\newblock Structural and practical identifiability analysis of partially
	observed dynamical models by exploiting the profile likelihood.
	\newblock {\em Bioinformatics}, 25(15):1923--1929, 2009.
	
	\bibitem{rey2022strikepy}
	D.~Rey~Rostro and A.~F. Villaverde.
	\newblock Strikepy: nonlinear observability analysis of inputs, states, and
	parameters in python.
	\newblock In {\em Actas de las XLIII Jornadas de Autom\'atica}, 2022.
	
	\bibitem{ritt1950differential}
	J.~F. Ritt.
	\newblock {\em Differential algebra}, volume~33.
	\newblock American Mathematical Soc., 1950.
	
	\bibitem{saccomani2001new}
	M.~P. Saccomani, S.~Audoly, G.~Bellu, and L.~D'Angio.
	\newblock A new differential algebra algorithm to test identifiability of
	nonlinear systems with given initial conditions.
	\newblock In {\em Proceedings of the 40th IEEE Conference on Decision and
		Control (Cat. No. 01CH37228)}, volume~4, pages 3108--3113. IEEE, 2001.
	
	\bibitem{saccomani2003parameter}
	M.~P. Saccomani, S.~Audoly, and L.~D'Angi{\`o}.
	\newblock Parameter identifiability of nonlinear systems: the role of initial
	conditions.
	\newblock {\em Automatica}, 39(4):619--632, 2003.
	
	\bibitem{sedoglavic2002probabilistic}
	A.~Sedoglavic.
	\newblock A probabilistic algorithm to test local algebraic observability in
	polynomial time.
	\newblock {\em Journal of Symbolic Computation}, 33(5):735--755, 2002.
	
	\bibitem{sedoglavic2007reduction}
	A.~Sedoglavic.
	\newblock Reduction of algebraic parametric systems by rectification of their
	affine expanded {Lie} symmetries.
	\newblock In {\em International Conference on Algebraic Biology}, pages
	277--291. Springer, 2007.
	
	\bibitem{shi2022efficient}
	X.~Shi and M.~Chatzis.
	\newblock An efficient algorithm to test the observability of rational
	nonlinear systems with unmeasured inputs.
	\newblock {\em Mechanical Systems and Signal Processing}, 165:108345, 2022.
	
	\bibitem{stigter2015fast}
	J.~D. Stigter and J.~Molenaar.
	\newblock A fast algorithm to assess local structural identifiability.
	\newblock {\em Automatica}, 58:118--124, 2015.
	
	\bibitem{thomas1989effect}
	G.~D. Thomas, M.~J. Chappell, P.~W. Dykes, D.~B. Ramsden, K.~R. Godfrey, J.~R.
	Ellis, and A.~R. Bradwell.
	\newblock Effect of dose, molecular size, affinity, and protein binding on
	tumor uptake of antibody or ligand: a biomathematical model.
	\newblock {\em Cancer research}, 49(12):3290--3296, 1989.
	
	\bibitem{topp2000model}
	B.~Topp, K.~Promislow, G.~Devries, R.~M. Miura, and D.~T~Finegood.
	\newblock A model of $\beta$-cell mass, insulin, and glucose kinetics: pathways
	to diabetes.
	\newblock {\em Journal of theoretical biology}, 206(4):605--619, 2000.
	
	\bibitem{tunali1987new}
	E.~Tunali and T.-J. Tarn.
	\newblock New results for identifiability of nonlinear systems.
	\newblock {\em IEEE Transactions on Automatic Control}, 32(2):146--154, 1987.
	
	\bibitem{verdiere2005identifiability}
	N.~Verdiere, L.~Denis-Vidual, G.~Joly-Blanchard, and D.~Domurado.
	\newblock Identifiability and estimation of pharmacokinetic parameters for the
	ligands of the macrophage mannose receptor.
	\newblock {\em International Journal of Applied Mathematics and Computer
		Science}, 15:517--526, 2005.
	
	\bibitem{villaverde2016structural}
	A.~F. Villaverde, A.~Barreiro, and A.~Papachristodoulou.
	\newblock Structural identifiability of dynamic systems biology models.
	\newblock {\em PLoS computational biology}, 12(10):e1005153, 2016.
	
	\bibitem{villaverde2018input}
	A.~F. Villaverde, N.~D. Evans, M.~J. Chappell, and J.~R. Banga.
	\newblock Input-dependent structural identifiability of nonlinear systems.
	\newblock {\em IEEE Control Systems Letters}, 3(2):272--277, 2018.
	
	\bibitem{villaverde2022protocol}
	A.~F. Villaverde, D.~Pathirana, F.~Fr{\"o}hlich, J.~Hasenauer, and J.~R. Banga.
	\newblock A protocol for dynamic model calibration.
	\newblock {\em Briefings in bioinformatics}, 23(1):bbab387, 2022.
	
	\bibitem{villaverde2019full}
	A.~F. Villaverde, N.~Tsiantis, and J.~R. Banga.
	\newblock Full observability and estimation of unknown inputs, states and
	parameters of nonlinear biological models.
	\newblock {\em Journal of the Royal Society Interface}, 16(156):20190043, 2019.
	
	\bibitem{walter1982global}
	E.~Walter and Y.~Lecourtier.
	\newblock Global approaches to identifiability testing for linear and nonlinear
	state space models.
	\newblock {\em Mathematics and Computers in Simulation}, 24(6):472--482, 1982.
	
	\bibitem{weber2001modeling}
	A.~Weber, M.~Weber, and P.~Milligan.
	\newblock Modeling epidemics caused by respiratory syncytial virus (rsv).
	\newblock {\em Mathematical biosciences}, 172(2):95--113, 2001.
	
	\bibitem{wieland2021structural}
	F.-G. Wieland, A.~L. Hauber, M.~Rosenblatt, C.~T{\"o}nsing, and J.~Timmer.
	\newblock On structural and practical identifiability.
	\newblock {\em Current Opinion in Systems Biology}, 25:60--69, 2021.
	
	\bibitem{wodarz1999specific}
	D.~Wodarz and M.~A. Nowak.
	\newblock Specific therapy regimes could lead to long-term immunological
	control of {HIV}.
	\newblock {\em Proceedings of the National Academy of Sciences},
	96(25):14464--14469, 1999.
	
	\bibitem{wolkenhauer2008parameter}
	O.~Wolkenhauer, P.~Wellstead, K.-H. Cho, J.~R. Banga, and E.~Balsa-Canto.
	\newblock Parameter estimation and optimal experimental design.
	\newblock {\em Essays in biochemistry}, 45:195--210, 2008.
	
\end{thebibliography}
\end{document}